\documentclass[letterpaper,twocolumn,superscriptaddress,floatfix]{revtex4}
\usepackage[latin1]{inputenc}
\usepackage{bm}
\usepackage{multirow,amssymb,amsbsy,amsmath}
\usepackage{graphicx}
\usepackage{verbatim}
\makeatletter
\usepackage{pifont}
\makeatother

\usepackage{graphicx}
\usepackage{dcolumn}
\usepackage{bm}
\usepackage{ifthen}
\usepackage{booktabs}
\usepackage{nicefrac}
\usepackage{float}
\usepackage{color,xcolor}

\begin{document}

\title{An ultrastable and robust single-photon emitter in hexagonal boron nitride}
\author{Wei Liu}
\author{Yi-Tao Wang}
\email{yitao@ustc.edu.cn}
\author{Zhi-Peng Li}
\author{Shang Yu}
\author{Zhi-Jin Ke}
\author{Yu Meng}
\author{Jian-Shun Tang}
\email{tjs@ustc.edu.cn}
\author{Chuan-Feng Li}
\email{cfli@ustc.edu.cn}
\author{Guang-Can Guo}
\affiliation{CAS Key Laboratory of Quantum Information, University of Science and Technology of China, Hefei, P.R.China}
\affiliation{CAS Center For Excellence in Quantum Information and Quantum Physics, University of Science and Technology of China, Hefei, P.R.China}

\date{\today }
\begin{abstract}
Quantum emitters in van der Waals (vdW) materials have attracted lots of attentions in recent years, and shown great potentials to be fabricated as quantum photonic nanodevices. Especially, the single photon emitter (SPE) in hexagonal boron nitride (hBN) emerges with the outstanding room-temperature quantum performances, whereas the ubiquitous blinking and bleaching restrict its practical applications and investigations critically. The blister in vdW materials possessing stable structure can modify the local bandgap by strains on nanoscale, which is supposed to have the ability to fix this photostability problem. Here we report a blister-induced high-purity SPE in hBN under ambient conditions showing stable quantum-emitting performances, and no evidence of blinking and bleaching for one year. Remarkably, we observe the nontrivial successive activating and quenching dynamical process of the fluorescent defects at the SPE region under low pressures for the first time, and the robust recoverability of the SPE after turning back to the atmospheric pressure. The pressure-tuned performance indicates the SPE origins from the lattice defect isolated and activated by the strain induced from the blister, and sheds lights on the future high-performance quantum sources based on hBN.

\end{abstract}


\maketitle 
\bibliographystyle{prsty}

 \begin{figure*}[t]
  \centering
  \includegraphics[width=0.95\textwidth]{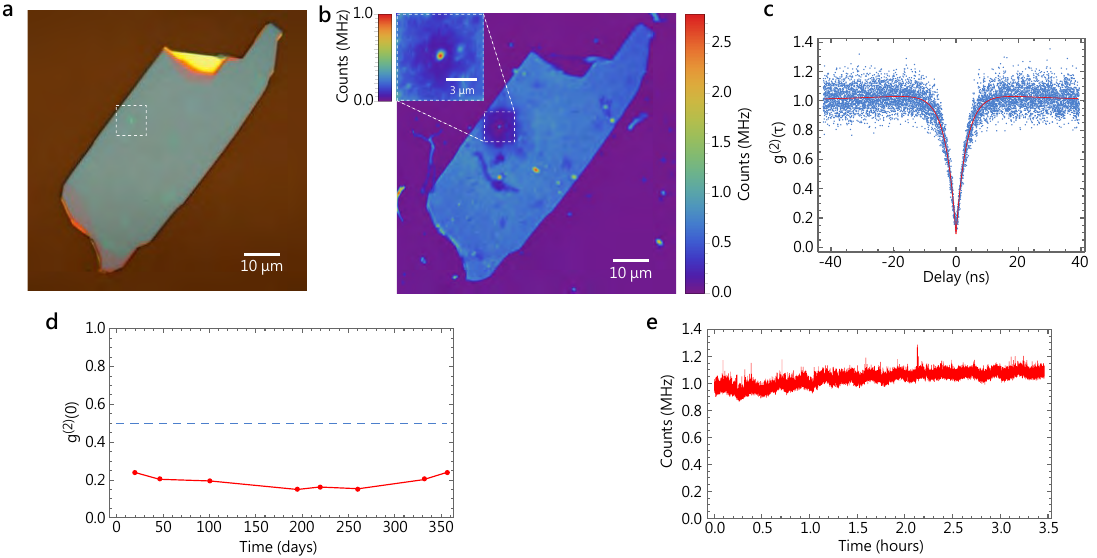}
  \caption{\textbf{The blister-induced blinking-free, bleaching-free and high-purity single-photon emitter (SPE) in the hexagonal boron nitride (hBN) flake.} (a) the optical microscope image of the hBN flake. (b) the confocal fluorescence image of the same hBN flake. The SPE is located at the center of the white dashed box marked in (a) and (b), enlarged by the top-left inset. (c) The antibunching second-order time correlation $g^{(2)}(\tau)$ of the SPE under 0.3-mW excitation with an additional long-pass filter (see Appendix B). The blue dots are the experimental data and the red curve shows the corresponding fitting result using three-level model. The lifetimes of the excited and metastable states are 3.38 ns and 24.95 ns, respectively. $g^{(2)}(0)=0.089  \pm 0.006$ indicates the SPE possesses the high single-photon purity. (d) The quantum-emitting performance of the SPE evaluated by $g^{(2)}(0)$ during one year. The blue dashed line indicates the quantum-emitting boundary of $g^{(2)}(0)=0.5$. No significant change of the blister profile is observed under ambient conditions during these days. (e) The photon counts of the SPE measured for continuing 3.5 hours, with the 100-ms time resolution. Both (d) and (e) illustrate the nearly perfect photostability of the blinking-free and bleaching-free SPE.}\label{fig1}
\end{figure*}

\section{Introduction}
Single-photon emitter (SPE) is the fundamental quantum resource in quantum-information technologies \cite{OBrien2009} such as linear optical quantum computation \cite{Kok2007}, quantum metrology \cite{Giovannetti2011} and quantum cryptography \cite{Gisin2002}. The SPEs in two-dimensional van der Waals (vdW) materials of transition metal dichalcogenides (TMDCs) and hexagonal boron nitride (hBN) have been discovered and attracted lots of attentions recently \cite{Tran2016,He2015,Srivastava2015}, on account for their unique heterogeneous-assembly \cite{Liu2016} and optoelectronics \cite{Schaibley2016,Xia2014} features, along with the great potentials to be integrated with the other nanophotonic devices to develop the high-performance hybrid quantum systems \cite{Aharonovich2016,Low2016,Basov2017}. For practical applications and investigations, the easy operation, tunability and photostability of SPEs are of great importance. HBN possesses the wide bandgap $\sim$6 eV in contrast with TMDCs \cite{Cassabois2016}, hence, can host the room-temperature SPE generated from the deep defects \cite{Tran2016}. The single fluorescent defects in hBN can be generated by various methods such as the ion or electron irradiations and chemical etching \cite{Choi2016,Chejanovsky2016}. Therefore, the room-temperature working ability and easy fabrication make SPEs in hBN take great advantages on the easy operation. Moreover, the SPEs in hBN associated with different crystallographic defects can contribute to photons at different wavelengths. The carbon substitutional impurity at a nitrogen site (C$_\text{N}$) in hBN could emit the ultraviolet single photons \cite{Bourrellier2016}, while the antisite nitrogen vacancy (N$_\text{B}$V$_\text{N}$) could generate the visible-wavelength \cite{Tran20162}. The lattice strains and defects circumstances in hBN also provide the effective tuning methods to modify the electronic structures and emitting wavelengths of the SPEs, leading to the wavelength-tunable broadband single-photon source based on hBN \cite{Tran20162,Grosso2017,Chejanovsky2016}.

The easy operation and various tunability make hBN an outstanding host system for the development of the solid-state quantum source. As the novel quantum emitter discovered recently, the SPE and related performances in hBN are still required to be study insightfully for the generation mechanism and intrinsic property. Therefore, the SPE is required to possess the essential property of the long-time photostability for various experimental investigations \cite{Vogl2018,Vogl2019}. However, the blinking \cite{Mart1nez2016,Tran20163,Choi2016,Shotan2016} and bleaching \cite{Chejanovsky2016,Exarhos2017,Exarhos2018}, which result in the serious photon-emitting fluctuation and even irreversible fluorescence quenching, are always ubiquitous for SPEs in hBN. Most SPEs we found and investigated in hBN flakes could keep quantum-emitting performances for only several hours or days.

Moreover, the defects concentration \cite{Chejanovsky2016} and background fluorescence \cite{Grosso2017} in hBN are also two obstacles to fabricate the high-purity SPEs in hBN efficiently. The former would generate various fluorescent points in hBN containing multiple emitters that makes it difficult to isolate the individual quantum emitters among the fluorescent points \cite{Chejanovsky2016}, or result in the defect-fluorescence quenching, e.g., via Auger recombination \cite{Kurzmann2016}; the latter contribute the major leading to the decrease of the single-photon-emitting purity \cite{Grosso2017}. These two shortcomings together with the photostability problem limit the practical applications and investigations of hBN-based SPEs, and become the roadblocks to these future high-performance quantum sources.

To solve these problems, the blister inside hBN, which can be utilized to implement the local strain engineering, is a good option. The blister containing bubble and tent in vdW materials has been employed as the powerful tool to study the unique mechanical properties \cite{Koenig2011,Khestanova2016,Lloyd2017}, interfacial metrology \cite{Wang2017,Dai2019}, piezoelectricity \cite{Ares2020}, and optoelectronics \cite{Harats2020,Hong2015} of vdW materials. Due to the strong adhesion and tough mechanical properties of vdW materials, the blister inside could maintain the stable structure and modify the materials performances on nanoscale, therefore, has also been used to realize the nanoscale engineering located at the blister and investigate many novel states of matters, such as the enormous magnetic fields \cite{Levy2010}, nanocrystal \cite{Yuk2012} and room-temperature ice \cite{Algara2015}. On account to the unique properties, the blister inside hBN could be utilized to improve the SPE performances. On one hand, the blister could provide the relatively freestanding surroundings and eliminate the affects from the substrate to improve the photostability of the SPE \cite{Exarhos2017,Exarhos2018}; On the other hand, the blister-induced strain could lead to the valley-shape or funnelled spatial energy band for vdW materials \cite{Wiktor2016,Feng2012}, and further enhance the fluorescence emitted from the bottom of the funnelled or valley-shape energy band \cite{Palacios2017,Branny2017,Shepard2017}. In addition, since the blister inside vdW materials could modify the strain and bandgap located at the blister region, it would possess the potential to develop the tunability (e.g., by the external pressure) on nanoscale. Therefore, the blister also provides a method to study the SPE behavior under the tunable bandgap by the blister-induced strain engineering.

Here, we report a robust blister-induced SPE in the hBN flake with the nearly perfect photostability for one year under the ambient conditions. The SPE shows no evidence of blinking, and maintains the stable quantum-emitting performances all the time. The background fluorescence from the surrounding of the SPE is much darker than the external region of the hBN flake, resulting in the high-purity single-photon emission. Interestingly, the SPE region shows a successive and repeatable activating and quenching dynamical process of the fluorescence and quantum-emitting performances at the low external pressure, which was not reported before and reflects the mechanical properties of hBN. The low-pressure performances corresponding to the activated fluorescent defect ensemble in hBN show little quantum-emitting performances but the enhanced fluorescence. The SPE also possesses the robust recoverability of the fluorescence and quantum performances after turning back to the ambient conditions. According to the comparison of the spectra at different pressures, it can be concluded that the blister inside hBN possesses the ability to activate the fluorescent defects by strains. The SPE is generated based on the single defect that is isolated and individually activated by the unexpanded blister (at the atmosphere pressure) from the defect ensemble. The investigation demonstrates the room-temperature, blinking-free, bleaching-free and high-purity SPE generated by the blister-induced strain in hBN, and reveals parts of photoluminescence mechanism of the individual defects in hBN.

\section{Results and discussion}

 \begin{figure}
  \centering
  \includegraphics[width=0.35\textwidth]{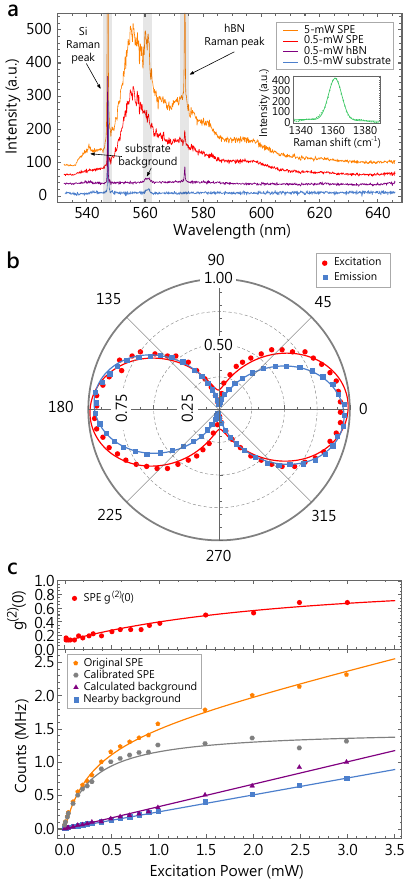}
  \caption{\textbf{The photoluminescence characters of the single-photon emitter (SPE).} (a) The fluorescence spectra of the SPE (orange and red curves, corresponding to the excitation power of 5 mW and 0.5 mW, respectively), hBN flake (purple curve) and the uncovered Si-SiO$_2$ substrate (blue curve). The raman peaks of Si and hBN, as well as the background fluorescence peaks from the substrate are labeled by the gray bands. The full view of the raman peaks at the 5-mW excitation are not shown since its too-high intensity. The Raman peak of the SPE is highlighted in the inset, where the green curve is the Gaussian-function-fitting result of the experimental data shown by the green dots. The raman shift of the SPE is $1361.18\pm 0.05$ cm$^{-1}$ with the FWHM of $9.80\pm0.14$ cm$^{-1}$. (b) The excitation (red dots and curve) and emission (blue squares and curve) polarization behaviors of the SPE. The curves are the corresponding cos$^2(\theta)$-fitting results where $\theta$ is the polarization angle. (c) the excitation-power-dependent $g^{(2)}(0)$ (red dots), original fluorescence intensities from the SPE (orange pentagons) and the hBN flake nearby the SPE (blue squares), calibrated fluorescence intensity from the SPE (gray hexagons) derived by eliminating the calculated background fluorescence intensity (purple triangles). The calculated background fluorescence intensity is derived from $g^{(2)}(0)$. The curves are the corresponding fitting results and theoretical curves (see Appendix C). }\label{fig2}
\end{figure}

\subsection{Stable quantum-emitting performances}

The optical and confocal fluorescence microscope images of the investigated hBN sample are shown in Fig. 1(a) and (b), respectively. The blister-induced SPE locates at the center of the white box marked in (a) and (b). The blister is generated in the hBN-transfer step during the sample preparation (see Appendix A). The background fluorescence around the SPE quenches remarkably compared with the other region of the hBN flake, and the intensity of that is as low as the fluorescence emitted from the Si-SiO$_2$ substrate as shown in Fig. 1(b). The SPE possesses the second-order time correlation $g^{(2)}(0)$ able to reach $0.089  \pm 0.006$ as shown in Fig. 1(c) derived from the three-level-model fitting result, corresponding to the single-photon purity of $\rho=(95.4\pm0.6)\%$ (see Appendix C). Besides the high single-photon purity, the SPE also possesses the ultra-long-time stable quantum-emitting performances under the ambient conditions as shown in Fig. 1(d). The measured $g^{(2)}(0)$ could keep $\sim$0.2 for one year without any postselection but a 532-nm long-pass filter to block the excitation laser. It is far below the oft-quoted quantum boundary of $g^{(2)}(0)=0.5$ all the time. The variations of $g^{(2)}(0)$ on these days are mainly resulted from the changes of the environmental background and the fluorescence-collection efficiency. The significant change of the blister profile is not observed under ambient conditions during these days. The SPE also keeps the long-time stable photon-emitting rate $\sim$1 MHz as shown in Fig. 1(e), where the rate approaches the saturation-radiation rate of the SPE under the 0.5-mW-excitation condition. The counts jitters at the time period $\sim$$1/4$ hours are mainly resulted from the jitters of the excitation laser and the optical elements.

\begin{figure*}[t]
  \centering
  \includegraphics[width=0.95\textwidth]{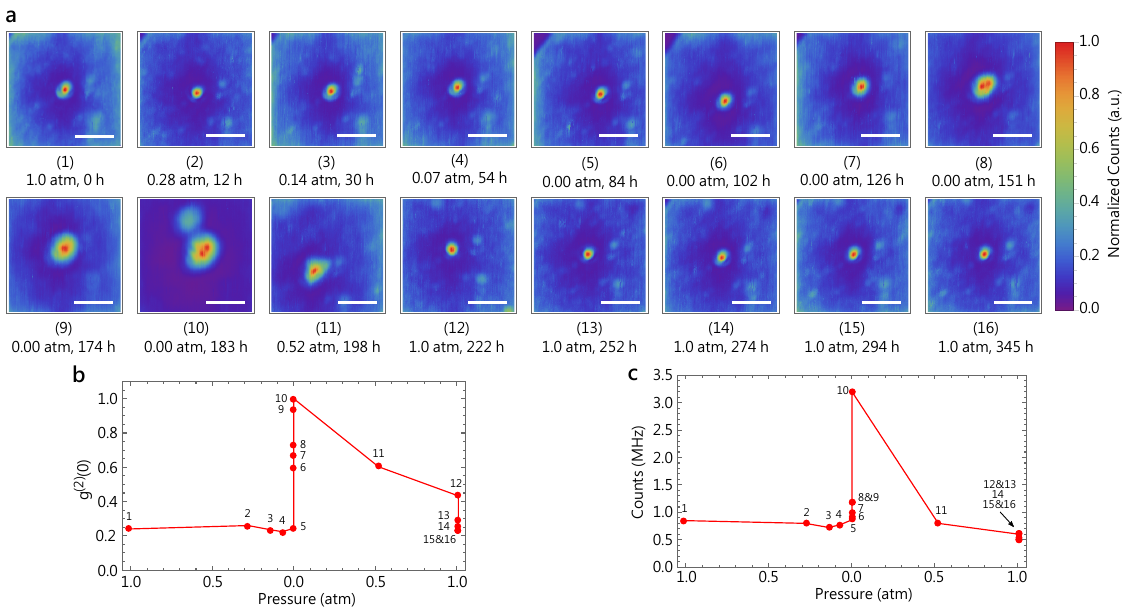}
  \caption{\textbf{The nontrivial photoluminescence variations of the single photon emitter (SPE) under low-pressure conditions.} (a) The confocal fluorescence images of the SPE region at different pressures over time (the below labels the marked number, external pressure, and the measuring moment of every image). The external pressure is changed to the next one as soon as the data acquisition finished, and decresed from the atmosphere pressure to 0.00 atm ($\sim$10$^{-2}$ Pa actually) and then increased back to the atmosphere pressure over time from state (1) to (16). Every image is normalized by the maximum fluorescence intensities of themselves to highlight the SPE profile changes. The white scale bars indicate 3 $\mu$m. (b)\&(c) The $g^{(2)}(0)$  and fluorescence intensities measured at the SPE position corresponding to the every state in (a) marked by the numbers 1$\sim$16. }\label{fig3}
\end{figure*}

The perfect photostability, i.e., bleaching-free and blinking-free property, should be related to the blister-induced freestanding surrounding at the SPE region, which eliminates the affects of the substrate and reduces the charge fluctuations, and provides the much inerter surrounding for the SPE \cite{Exarhos2017,Exarhos2018,Tonndorf2015}. As a comparison, the suspended SPE in hBN flake reported in Ref. \cite{Exarhos2017} can also be stable in several months under low-power excitation; but the high-power (i.e., over $\sim$0.15 mW) excitation causes the SPE to bleach away either. This bleaching is perhaps resulted from the lattice damage induced by the inhomogeneous heating from the focused excitation laser on the large-area freestanding hBN flake ($\sim$5 $\mu$m). However, the blister-induced SPE maintains nearly perfect photostability under the high-power (even up to 10-mW) excitation. It should be related to the tiny and relatively steady freestanding structure of the blister inside.

\subsection{Photoluminescence characters}

Figure 2 shows the photoluminescence characters of the SPE. The zero-phonon line (ZPL) and two phonon-side bands (PSBs) can be estimated by the three-Gaussian-function fitting on the 0.5-mW-excitation spectrum in the photon-frequency domain. The fitting results indicate that the ZPL centers $\sim$556 nm (FWHM=8.48 nm), as well as two PSBs center $\sim$566 nm (FWHW=28.1 nm) and $\sim$596 nm (FWHM=18.7 nm). The background peak centering $\sim$561 nm marked by the gray band in Fig. 2(a)  emerges from the background of the Si-SiO$_2$ substrate (but not from the hBN flake), and contributes the major to the background fluorescence. The excitation and emission polarization characters are shown in Fig. 2(b), with the derived degrees of polarization (DOPs) of 84.8\% and 96.4\%, respectively (see Appendix D). The high emission DOP in good agreement with the single-photon purity $\rho$ indicates that the SPE is perhaps generated by a high-purity and stable electric-dipole-like defect. Figure 2(c) shows the excitation-power-dependent $g^{(2)}(0)$ and related fluorescence intensities of the SPE and background. The saturation-radiation rate of the SPE is derived as 1.49 MHz, as well as the saturating-excitation power as 0.28 mW (see Appendix C). The calculated background fluorescence at the SPE is slightly stronger than the nearby fluorescence, perhaps resulted from the fluorescence enhancement at the point of the SPE.

\subsection{Dynamical defect-activating and -quenching processes}

The investigation above shows the elementary features of the SPE at the atmosphere pressure. The particularities of the SPE include: (1) the dark surrounding fluorescence and high-purity single-photon emission, (2) the bleaching-free and blinking-free photostability, and (3) the high DOP. Under the low-pressure conditions, significant photoluminescence changes appear at the SPE region. It reveals that the generation mechanism of the SPE is closely related to the blister-induced activating and isolating effects for the defects in the hBN flake treated by the high-density ion irradiation (see Appendix E).

The photoluminescence performances of the SPE at different pressures over time are shown in Fig. 3. Since the pressure change does not influence the fluorescence instantly, the SPE is hold under every pressure condition for more than ten hours to investigate the pressure-dependent fluorescence. The nontrivial performances of the low-pressure fluorescence appear as followings:

(i) For the state (6) in Fig. 3, $g^{(2)}(0)$ starts increasing significatively from 0.24 to 0.60, while the fluorescence intensity of the SPE remains nearly unchanged. The changed $g^{(2)}(0)$  should be related to the reduction of the single photons emitted from the SPE, and the enhancement of the background fluorescence.

(ii) For the state (7)$\sim$(9) in Fig. 3, the profile of the SPE extends gradually along with the increase of the duration time. It indicates that the fluorescence-enhanced region extends from the original point. The $g^{(2)}(0)$ and fluorescence intensities also increase gradually.

(iii) Especially, for the state (10) in Fig. 3, the SPE shows the sudden-enhanced fluorescence $\sim$3 times of the original state. $g^{(2)}(0)$ reaches $\sim$1 corresponding to the uncorrelated fluorescence without quantum-emitting performances.

(iv) For the state (11)$\sim$(16) in Fig. 3, the $g^{(2)}(0)$ and fluorescence intensity both decrease. After the SPE is brought back to the atmosphere conditions and relaxed for days, the SPE in the state (16) recovers to the same as the origin state (1). The performance (iv) demonstrates the robustness of the SPE, i.e., the perfect quantum performances recoverability even after that is entirely destroyed.

\begin{figure}[t]
  \centering
  \includegraphics[width=0.35\textwidth]{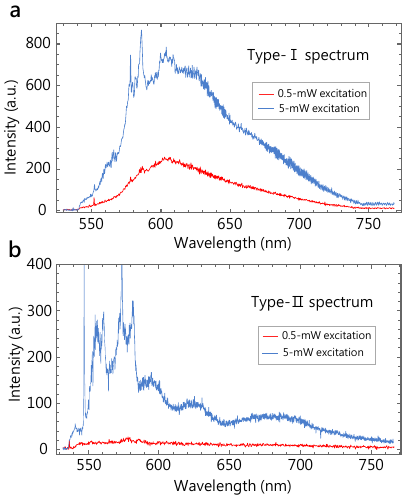}
  \caption{\textbf{Two typical photoluminescence spectra at the region of the single photon emitter induced by the expanded blister under low-pressure conditions.} (a) The type-\uppercase\expandafter{\romannumeral1} spectra corresponding to the large group of defect ensemble activated. (b) The type-\uppercase\expandafter{\romannumeral2} spectra corresponding to the small group of defect ensemble activated. The red and blue curves in (a) and (b) are acquired under 0.5-mW and 5-mW excitation conditions, respectively.}\label{fig2}
\end{figure}

The above nontrivial performances could be related to the pressure-dependent deformation of the blister inside the hBN flake at the SPE region. Since the profile of the spatial strain applied on the vdW membrane can be directly mapped to the spatial-bandgap profile, the funnelled or valley-shape spatial bandgap can be generated from the strain induced by the blister \cite{Feng2012,Hong2015,Johari2012,Roldan2015}. The profile of the blister and related deformation strain expand with the decrease of external pressures, therefore, the strain-induced enhanced fluorescence at the SPE region would be tuned by the external pressure. It should be noted that the photoluminescence response takes multiple hours under the external-pressures tuning as shown in Fig. 3(a), which may be related to the slow diffusion of the molecules trapped in the blister. The trapped molecules would diffuse among the multiple layers of hBN flake, and lead to the slow expansion of blister under the low pressure.

As the vdW membrane, hBN flake possesses the strain-tunable bandgap \cite{Grosso2017,Wiktor2016}. The total strain distributed on the most blister region is in negative correlations with its distance from the blister center. On account that the bandgap of hBN possesses the negative deformation potentials with strain, i.e., $\sim$-31 meV per 1\% strain \cite{Wiktor2016}, the blister inside hBN leads to the funnelled or valley-shape bandgap and further concentrated excitons at the region of blister (see Appendix E for the details of the theoretical calculation for the spatial distributions of the blister-induced strain and the related bandgap for the tent and bubble models in hBN). The excitons near the blister region would possess the higher probability to tunnel or flow to the bottom of the bandgap, as well as the lower bandgap would lead to the higher-density charge carriers according to the Boltzmann distribution $\rho=\rho_0e^{-\frac{E}{kT}}$ (with ground-state carrier density $\rho_0$, carrier energy $E$, temperature $T$ and Boltzmann constant $k$). Therefore, the blister-induced strain gives rise to the concentrating of the excitons into the blister center and the locally enhanced fluorescence in the hBN flake. It induces that the SPE emerges with the performances (i) and (ii), i.e., the remarkably increased $g^{(2)}(0)$ and the slightly enhanced fluorescence intensity along with the expanding of the strain-induced bandgap in the hBN flake.

However, the performance (iii) emerges with the remarkable differences compared with the other states on the fluorescence intensities and fluorescent profiles. The performance (iii) possesses the seriously- and suddenly-enhanced fluorescence, as well as the much larger fluorescence-enhanced region. The fluorescence enhancement mechanism in the performance (ii) can be simply explained by the increase of the charge-carriers density around the SPE, while that in the performance (iii) should be different. It implies that the blister should possess the more powerful ability, i.e., to activate the photoluminescence of defects, besides the local-fluorescence enhancement and photostability. Two typical photoluminescence spectra at the SPE region under low-pressure conditions are shown in Fig. 4 marked by type-\uppercase\expandafter{\romannumeral1} and type-\uppercase\expandafter{\romannumeral2}. Type-\uppercase\expandafter{\romannumeral1} spectra correspond to the special state (10), and type-\uppercase\expandafter{\romannumeral2} spectra are the typical ones under the other low-pressure conditions corresponding to the states (6)$\sim$(9) \& (11).

Compared with the spectra under the atmosphere conditions shown in Fig. 2(a) (i.e., the spectra of the high-purity SPE), both type-\uppercase\expandafter{\romannumeral1} and  type-\uppercase\expandafter{\romannumeral2} spectra exhibit the significantly broadened envelopes, corresponding to the fluorescence emitted from the defect ensemble in hBN at the SPE region. Type-\uppercase\expandafter{\romannumeral1} spectrum exhibits the nearly uniform broadened envelope, while type-\uppercase\expandafter{\romannumeral2} spectrum emerges with many peaks, especially under the high-power (i.e., 5-mW) excitation. It indicates that there should be a large group of defects activated by the expanded blister in state (10), while just a small group of defects activated in the other low-pressure states.

\section{Conclusions}

The comparation of the high-purity-SPE, type-\uppercase\expandafter{\romannumeral1} and type-\uppercase\expandafter{\romannumeral2} spectra indicates that the blister-induced strain provides the significant fluorescence activating effect for the defects in hBN. There are always a group of defects at the SPE region in our investigations, while most of them stay at the dark states under the ambient conditions. Quantized energy levels related to the bright and dark states perhaps exist in these defects \cite{Kianinia2018}. The energy band locally modified by the blister may lead the defects transiting from the dark state to the bright level and activate the defects. Therefore, the SPE region exhibits the defect-ensemble-photoluminescence performances induced by the expanded blister under the low pressure. Whereas under the atmosphere conditions, the unexpanded small-size blister activates just one individual defect, therefore, could generate the high-purity SPE (i.e., isolate the single defect) among the defect ensemble. The detailed study is still required for the defect level structure and the related strain-modified photon physics.

It should be point out that the high-dose ion irradiation perhaps contributes to the dark-defect-ensemble generation in the sample-preparation process (i.e., $10^{14}$ ions cm$^{-2}$, see Appendix A) \cite{Chejanovsky2016}. We also experimentally investigate the hBN flakes treated with the normal dose of $10^{10}$ cm$^{-2}$, which is the usual dose to generate the single defects in hBN \cite{Choi2016}. The number of SPEs exhibiting the antibunching behaviors in the normal-dose-irradiation hBN flakes is much larger than that in the high-dose-irradiation hBN flakes. It indicates that the high-dose nitrogen-ion irradiation is related to the defect quenching in hBN, on account that the generated-defect number should be in the positive correlation with the irradiation dose. The quench is perhaps resulted from the excessive concentration of the generated defects and the further intensified Auger recombination or lattice damages.

In conclusion, we experimentally investigate the blister-induced SPE in the hBN flake. Due to the steady structure and freestanding surrounding of the blister inside, the SPE possesses the outstanding advantages of the high single-photon purity, bleaching-free and blinking-free photostability under the ambient conditions for one year all the time. Besides, the successive dynamical process of the activating and quenching of the individual defects are observed and investigated under low-pressure conditions for the first time. The results indicate the powerful abilities of the blister to isolate and activate the single defects in hBN, among the nonradioactive defect ensemble generated from the high-dose ion irradiation. This work reveals the SPE origination mechanism in hBN via the blister inside, and sheds lights on the fabrication for the room-temperature high-performance quantum source with ultrastability and robustness, moreover, exhibits the remarkable potential of the tunable blister-induced quantum emitter in vdW materials for the investigations of the related defects property.

\section*{Funding Information}

This work is supported by the National Key Research and Development Program of China (No. 2017YFA0304100), the National Natural Science Foundation of China (Grants Nos. 61327901, 11674304, 11822408, 61490711, 11774335, 11821404, and 11904356), the Key Research Program of Frontier Sciences of the Chinese Academy of Sciences (Grant No. QYZDY-SSW-SLH003), the Youth Innovation Promotion Association of Chinese Academy of Sciences (Grants No. 2017492), the Foundation for Scientific Instrument and Equipment Development of Chinese Academy of Sciences (No. YJKYYQ20170032), Science Foundation of the CAS (No. ZDRW-XH-2019-1), Anhui Initiative in Quantum Information Technologies (AHY020100, AHY060300), the National Postdoctoral Program for Innovative Talents (Grant No. BX20180293), China Postdoctoral Science Foundation (Grant No. 2018M640587), the Fundamental Research Funds for the Central Universities (No. WK2470000026 and WK2030000008).

\section*{Acknowledgments}

The authors thank Wei-Ping Zhang, Xiong Zhou and Li-Ping Guo for the technical support on the ion implantation experiments carried out on the implanter (JZM 5900) at the Acelerator Laboratory of Wuhan University. This work was partially carried out at the USTC Center for Micro and Nanoscale Research and Fabrication.

\section{APPENDIX A: Sample preparation}
The hexagonal boron nitride (hBN) sample is mechanically exfoliated from the bulk-crystal hBN (HQ Graphene) and prepared as the multilayer flake onto the silicon (Si) wafer with the 300-nm SiO$_2$ toplayer. The Si-SiO$_2$ substrates are coated with Au marks as the location coordinates in the microscope, prepared by the lift-off technology based on the ultraviolet lithography. The substrates are cleaned by acetone, isopropyl alcohol and deionized water successively before using. The prepared hBN sample is irradiated by 3-keV nitrogen ions with the dose of $10^{14}$ cm$^{-2}$ to generate the defects, followed by the anneal treatment at 850 $^\circ$C under the 0.5-torr argon atmosphere for 30 minutes. After that, the hBN sample is transferred to another Si-SiO$_2$ substrate and eliminate the background fluorescence emitted from the residual organics on the original substrate, which is enhanced after the high-temperature anneal. The hBN transfer is applied by the following steps: (i) attaching and pressing the polydimethylsiloxane (PDMS) film (GelPak, PF-30-X4) on the original substrate, (ii) peeling off the PDMS film rapidly to pick up the hBN flakes, (iii) attaching and pressing the PDMS film on another cleaned substrate, (iv) peeling off the PDMS film slowly and leave the hBN flakes on the substrate. The blister could be generated from the lift of some nanoparticle below or the aggregation of some trapped contents like water and hydrocarbons during the step (iii) and (iv) \cite{Ares2020,Dai2018}.

\section{APPENDIX B: Optical characterization}
The prepared hBN sample is characterized by the home-made confocal-microscope system. A 532-nm linear-polarized continue-wave (CW) laser is reflected by the 532-nm dichroic mirror, with the following galvanometer mirror to adjust the propagating angle. The following 4-f system consist of two lenses with the focal length f=10 cm, is used to map the plane-light field at galvanometer mirror to the entrance pupil of the NA=0.9 objective (Olympus MPLFLN100xBDP). The excitation laser is focused onto the hBN sample by the objective and scans on the sample by adjusting the propagating angle. The emitted fluorescence from the sample is collected by the objective and sent to the dichroic mirror, then filtered by the 532-nm long-pass filter to block the excitation laser and coupled into the 630-nm single-mode fibre. In the case to verify the single-photon purity of $\rho=95.4$\% (i.e., $g^{(2)}(0)=0.089$), an additional 563-nm long-pass filter is added to block the dominated background fluorescence generated from the Si-SiO$_2$ substrate (but not from the hBN flake). Considering the 563-nm long-pass filter also blocks portions of the photons emitted from the SPE, only the 532-nm long-pass filter is used for the other investigation cases, and the resulted $g^{(2)}(0)$ is always above 0.13 affected by the unblocked background fluorescence from the substrate. The coupled fluorescence is sent to the spectrometer to obtain the fluorescence spectra, or is split by another 50:50 fiber coupler and detected by two following single-photon avalanche diodes (SPADs), which constitute the Hanbury-Brown-Twiss (HBT)  interferometer. The signals from the SPADs are analysed by the time-to-digital converter and related programs to obtain the fluorescence intensity and the second-order time correlation $g^{(2)}(\tau)$. To investigate the fluorescence performances at different pressures, the objective and hBN sample are placed in the air-tightness chamber with an optical window. The pressure is adjusted by the air pump connected to the chamber.

\section{APPENDIX C: Second-order time correlation and excitation-power-dependent fluorescence}
The single-photon purity $\rho$ can be defined as $\rho=I_S/(I_S+I_B)$, where $I_S$ is the single-photon intensity emitted from the single-photon emitter (SPE), and $I_B$ is the fluorescence intensity from the background. The second-order time correlation of the background fluorescence is assumed as 1, corresponding to the uncorrelated Poisson-statistics fluorescence. The measured second-order time correlation $g^{(2)}(\tau)$  can be derived as \cite{Brouri2000}
\begin{equation}\label{equm1}
g^{(2)}(\tau)=1-\rho^2+\rho^2 g^{(2)}_S(\tau),
\end{equation}
where $\tau$ is the delay time in the HBT interferometer, and $g^{(2)}_S(\tau)$ is the ideal second-order time correlation from the SPE, which equals to 0 when $\tau=0$. Therefore, the single-photon purity of the SPE can be estimated by $\rho=\sqrt{1-g^{(2)}(0)}$. In the experiment, the second-order time correlation is derived from the fitting results of the coincidence counts $G^{(2)}(\tau)$ of the HBT interferometer. The used fitting function for the three-level model is
\begin{equation}\label{equm2}
G^{(2)}(\tau-\tau_0)=N\{1-\rho^2[(1+a)e^{-|\tau-\tau_0|/t_1}-ae^{-|\tau-\tau_0|/t_2}]\},
\end{equation}
where the fitting parameters include the effective lifetimes of the excited and metastable states $t_1$ and $t_2$, respectively, constant delay time $\tau_0$, single-photon purity $\rho$, population weight $a$, and normalized coefficient $N$. The measured second-order time correlation is derived by $g^{(2)}(\tau)=G^{(2)}(\tau)/N$. The intensity of the measured raw SPE fluorescence ($I_T$) includes the single photons from the SPE ($I_S$) and the background fluorescence ($I_B$), i.e., $I_T=I_S+I_B$. According to the measured second-order time correlation $g^{(2)}_S(0)=1-\rho^2$, the calibrated single-photons intensity from the SPE can be derived by $I_S=\sqrt{1-g^{(2)}_S(0)}I_T $, and fit by the function $I_1(P)=c_1P/(c_2+P)$, where $c_1$ and $c_2$ are the fitting parameters, and $P$ is the excitation power. $c_1$ equals to the saturation-radiation rate, and $c_2$ is the saturation-excitation power of the SPE. The calculated background fluorescence can be derived by $I_B=(1-\sqrt{1-g^{(2)}_S(0)})I_T$, and fitted by the linear function $I_2(P)=d_1P+d_2$, where $d_1$ and $d_2$ are the fitting parameters. Furthermore, the theoretical curve of the measured original SPE fluorescence can be derived by the above fitting results $I_1+I_2$, also the theoretical curve of $g^{(2)}(0)$ can be derived by $1-[I_1/(I_1+I_2)]^2$.

\section{APPENDIX D: Polarization \& spectrum fitting setups}
The function used to fit the polarization behavior of excitation or emission is $D(\theta)=e_1+e_2\cos^2{(\theta+e_3)}$, where $e_1$, $e_2$ and $e_3$ are the fitting parameters, and $\theta$ is the polarization angle of the excitation laser or emitted fluorescence. The degree of polarization (DOP) is derived by $e_2/(2e_1+e_2)$, corresponding to the visibility of $(D_\text{max}-D_\text{min})/(D_\text{max}+D_\text{min})$. The three-Gaussian function used to fit the photoluminescence spectrum of SPE is
\begin{equation}\label{equm3}
I(\nu)=\sum^{3}_{n=1}M_n\exp{\left\{-\left[\frac{\nu-\nu_n}{w_n/(2\sqrt{\text{In}2)}}\right]^2\right\}},
\end{equation}
where $\nu$ is the photon frequency, and the fitting parameters include the normalized coefficient $M_n$, center frequency $\nu_n$, and frequency-domain FWHM $w_n$ of the zero phonon line (ZPL) and two phonon sidebands (PSB)  (n=1,2,3).

\section{APPENDIX E: Strain distributions}

\begin{figure}
  \centering
  \includegraphics[width=0.45\textwidth]{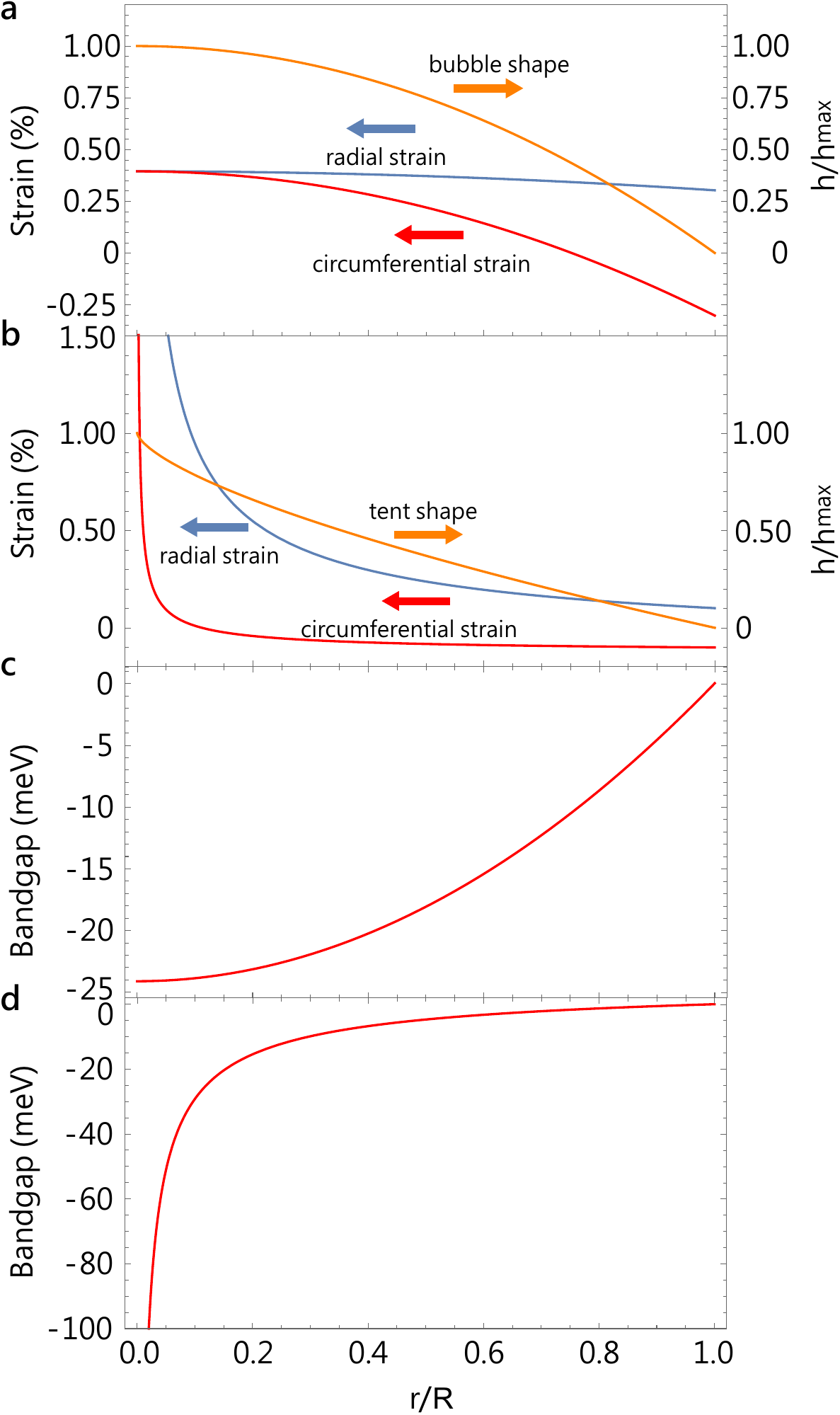}
  \renewcommand{\thefigure}{S1}
  \caption{\textbf{The strains and bandgap distributions on blister for bubble model and tent model.}
   (a) and (b) The blue and red curves indicate the radial and circumferential strains induced by the (a) bubble and (b) tent, respectively, corresponding to the left y-axis. The orange curve indicates the (a) bubble and (b) tent shape in the Hencky mode corresponding to the right y-axis. (c) and (d) The calculated spatial bandgap shift induced by the (c) bubble and (d) tent.}\label{figs1}
\end{figure}

Treating the hBN flake as a thin elastic membrane, and considering an axisymmetric round blister with the maximal radius $R$ and the maximal height $h_{\text{max}}$ inside the hBN flake, the shape of the blister can be described by the Hencky model
\begin{equation}
\frac{h}{h_{\text{max}}}=1-\left(\frac{r}{R}\right)^\alpha,
\end{equation}
where $r$ is the horizontal distance from the blister center, $h$ is the blister height, and $\alpha$ equals to 2/3 for the tent model and 2 for the bubble model. Then the radial and circumferential strains on the blister can be derived as \cite{Dai2018}:
\begin{equation}
\varepsilon_r(r)=c_1\left(\frac{h_{\text{max}}}{R}\right)^2\left[ 1-c_2\left(\frac{r}{R}\right)^{2\alpha-\color{blue}2\color{black}}\right]-c_3\left(\frac{h_{\text{max}}}{R}\right)^2,
\end{equation}
\begin{equation}
\varepsilon_{\theta}(r)=c_1\left(\frac{h_{\text{max}}}{R}\right)^2\left[ 1-\left(\frac{r}{R}\right)^{2\alpha-\color{blue}2\color{black}}\right]-c_3\left(\frac{h_{\text{max}}}{R}\right)^2,
\end{equation}
with
\begin{equation}
c_1=\frac{\alpha(2\alpha-1-\nu)}{8(\alpha-1)},
c_2=\frac{1+\nu-2\alpha\nu}{2\alpha-1-\nu},
c_3=\frac{\alpha(1+\nu)}{8}
\end{equation}
where $\nu$ is the Poisson's ratio of the membrane. For the bubble model, on assumption that the shape of the bubble inside the hBN flake possesses the similar behavior as that inside the monolayer hBN, the condition can be used that $h_{\text{max}}/R\approx l$ where $l$ is a constant independent with the bubble size \cite{Khestanova2016}. We use the Poisson's ratio of 0.211 \cite{Falin2017}, and $l\approx0.1$ referring to the monolayer hBN \cite{Khestanova2016} to estimate the hBN flake. For the tent model, the value of $l$ is also taken as $0.1$ for example. The calculated strain distributions on the blister for bubble model and tent model are shown in Fig. S1(a) and (b).

Furthermore, the blister-induced strain-tunable bandgap of hBN can be estimated from the deformation potentials $a_{rr}$ and $a_{\theta\theta}$ defined as
 \begin{equation}
a_{rr}=\frac{\Delta E_{rr}(r)}{\varepsilon_r(r)} \quad\text{and}\quad  a_{\theta\theta}=\frac{\Delta E_{\theta\theta}(r)}{\varepsilon_{\theta}(r)},
\end{equation}
where the total energy shift is derived by $\Delta E(r)=\Delta E_{rr}(r)+\Delta E_{\theta\theta}(r)$ under the approximate condition that the nonlinear deformation potential $a_{r\theta}\approx0$. According to the data provided in Ref. \cite{Wiktor2016}, two uniaxial deformation potentials along the perpendicular directions in hBN are given by $a^{\text{gap}}_{xx}=-3.05$ eV and $a^{\text{gap}}_{yy}=-3.06$ eV. Considering the case with $a_{rr}=a^{\text{gap}}_{xx}$ and
$a_{\theta\theta}=a^{\text{gap}}_{yy}$ as an example, the calculated spatial bandgap shift $\Delta E(r)=a^{\text{gap}}_{xx}\varepsilon_r(r)+a^{\text{gap}}_{yy}\varepsilon_{\theta}(r)$ is shown in Fig. S1(b) and (c).
The bubble inside hBN could induce the strain-tunable valley-shape bandpap with the depth $\sim$-24 meV at the bottom, i.e., the center of the bubble. The tent inside hBN could induce the strain-tunable funnelled bandpap. Therefore, the blister inside could lead to the concentrated charge carriers and intensified photoluminescence at the blister region in hBN.

\section{APPENDIX F: Estimation of the confinement pressure}

\begin{figure}
  \centering
  \includegraphics[width=0.45\textwidth]{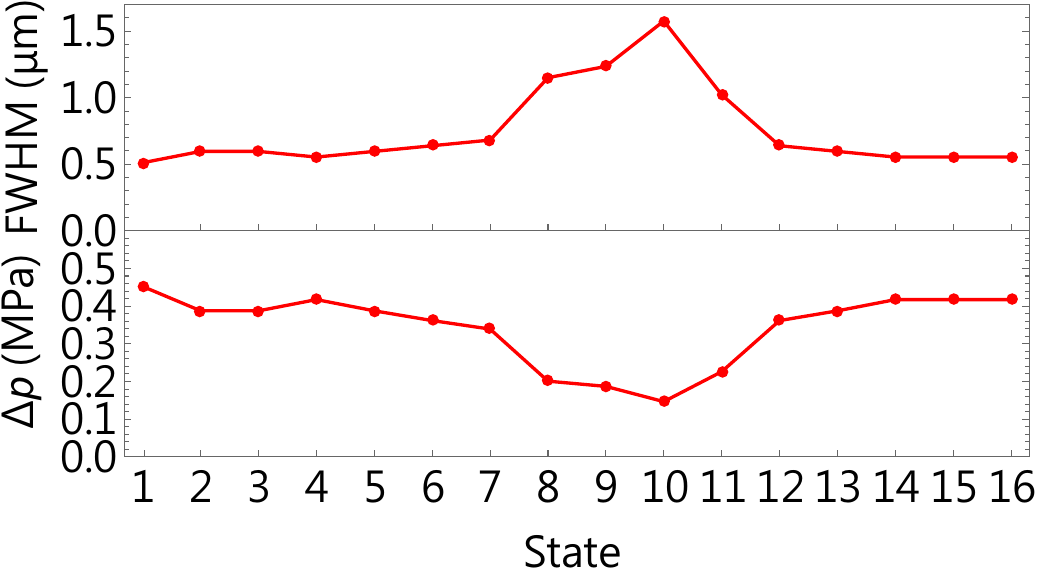}
  \renewcommand{\thefigure}{S2}
  \caption{\textbf{The full width at half maximum (FWHM) of SPE region and estimated confinement pressure $\Delta p$ in blister for state (1)$\sim$(16) shown in Fig. 3(a)}}\label{figs1}
\end{figure}

The confinement pressure in the blister can be estimated by the relation \cite{Sanchez2018}
 \begin{equation}
\Delta p=\frac{1}{R}\left(\eta E_{\text{2D}}l^3 \right),
\end{equation}
where $\eta\approx 1.6$ for the weak shear interface and in-plane elastic stiffness $E_{\text{2D}}=289$ N/m for hBN. To estimate the varying pressure in the blister of state (1)$\sim$(16) shown in Fig. 3(a), the blister radius $R$ is estimated by the half width of SPE region, and the aspect ratio $l$  is taken as 0.1 for example. The full width at half maximum (FWHM) of the fluorescence at SPE region and the estimated confinement pressure $\Delta p$ for state (1)$\sim$(16) are shown in Fig. s2, where $R$ equals to the half of FWHM. For state (5)$\sim$(10), the estimated pressure changes from 0.39 MPa to 0.15 MPa, with the outside pressure keeping at 0 atm.

\end{document}